\documentclass[conference]{IEEEtran}
\IEEEoverridecommandlockouts
\usepackage[T1]{fontenc}
\usepackage[utf8]{inputenc}
\usepackage{cite}
\usepackage{amsmath,amssymb,amsfonts}
\usepackage{algorithmic}
\usepackage{graphicx}
\usepackage{tabularx}
\usepackage{textcomp}
\usepackage{xcolor}
\usepackage{array}
\usepackage{booktabs}
\usepackage{url}
\usepackage{dblfloatfix}
\usepackage[none]{hyphenat}

\def\BibTeX{{\rm B\kern-.05em{\sc i\kern-.025em b}\kern-.08em
    T\kern-.1667em\lower.7ex\hbox{E}\kern-.125emX}}
\begin{document}

\title{Agentic Metaverse Services: A New As-a-Service Paradigm
\thanks{\IEEEauthorrefmark{1} Xiaofei Xu is the corresponding author.}
}

\author{
  \IEEEauthorblockN{
    Xiaofei Xu\textsuperscript{1*},
    Quan Z. Sheng\textsuperscript{2},
    Zhongjie Wang\textsuperscript{1},
    Boualem Benatallah\textsuperscript{3},
    Xiao Wang\textsuperscript{1},
    Ruipeng Han\textsuperscript{1}
  }
  \IEEEauthorblockA{
    \textsuperscript{1}Faculty of Computing, Harbin Institute of Technology, Harbin, China \\
    \textsuperscript{2}School of Computing, Macquarie University, Sydney, Australia \\
    \textsuperscript{3}School of Computing, Dublin City University, Dublin, Ireland \\
    \{xiaofei, rainy, wxlxq\}@hit.edu.cn; michael.sheng@mq.edu.au; boualem.benatallah@dcu.ie; ruipenghan@stu.hit.edu.cn\\
  }
}

\maketitle

\begin{abstract}
Generative Artificial Intelligence (GenAI) is reconstructing the digital virtual world, upgrading 
%the 
agents through enhancing 
%the 
their 
abilities 
%of agents 
in autonomous learning, multi-modal interaction, content generation, and collaborative decision-making. 
In particular, the shift from conversational chatbots to agentic AI, the most recent significant technical breakthrough of GenAI, has brought a new form of services, agentic services and Agent-as-a-Service (AaaS), in which the agent’s abilities are encapsulated, such as perception, decision-making, execution, collaboration, and content generation, to provide the customized agent services to users. The metaverse is a virtual ecosystem for human life, work, creation, and entertainment, supported by the new generation of digital technologies. 
Through combining agentic services and the metaverse, 
%the 
an 
Agentic Metaverse Service, denoted as AMServ, is produced for metaverse business processing, as a new form of metaverse service. The AaaS in the metaverse environment, denoted as Meta-AaaS, as an approach to realize AMServ, has become a new paradigm of agentic services and service computing. This paper overviews the evolution and new features of agents and services empowered by GenAI, reveals the roles and principles of agentic services in the metaverse environment, presents the forms, characteristics, and principles of the AMServ and the Meta-AaaS, discusses the typical application examples of the AMServ and the Meta-AaaS, and finally points out the new tendencies and research directions of the AMServ and the Meta-AaaS. The AMServ and the Meta-AaaS will bring great opportunities to human society and services in the AI era, and promote the rapid development of emerging service industries in the future.
\end{abstract}


% Alternative title from the source document:
% Or Metaverse AaaS -- A New Form of Agentic Services

\section{Introduction}

Currently, the agentic AI is leading a new interesting field of Generative Artificial Intelligence (GenAI), which has a big influence on service computing and related application domains. GenAI empowered agents with enhanced abilities, e.g., autonomous learning, multimodal interaction, content generation, collaborative decision-making, and proactive action, will improve the development and operation of services in many aspects. As the cross-field of agentic AI, metaverse and service computing, the Agentic Metaverse Services, denoted as AMServ, become a new form of services, and bring new challenges of services in the metaverse environment~\cite{xu2023metaverse, nisiotis2025emerging}. To understand and conduct research on the principles and performance of Agentic Metaverse Services, it is necessary to analyze the technical development trajectories of the related technologies, e.g., agents, service computing, cloud computing, and the metaverse. 
The findings and insights presented in the paper are derived from synthesizing related literature,  analyzing the existing technologies, our research experience, and the industrial case studies.
To realize AMServ in the physical-virtual fused space, a new paradigm, the Agent-as-a-Service (AaaS) in the metaverse, denoted as Meta-AaaS, is presented in this paper. As a new “as-a-Service” paradigm, the Meta-AaaS 
%can be seen as 
is 
an extension of SaaS (Software-as-a-Service) or XaaS (Everything-as-a-Service) in cloud computing, and an implementation form of agentic services in the metaverse environment. With more and more application scenarios of the AMServ and the Meta-AaaS into people’s life and business ecosystems, open challenging problems should be solved with new theories, methods, and technical approaches.

% In this visionary paper, we will overview the background and evolution trajectories of agents, services, XaaS, and the combination of agentic AI with metaverse services, reveal the roles and principles of agentic services in the metaverse environment, present the forms, characteristics, architecture and principles of the AMServ and the Meta-AaaS as a new “as-a-Service” paradigm, show the effects of the typical application examples of the AMServ and the Meta-AaaS, and finally discuss several important new research topics of the AMServ and the Meta-AaaS.

In this visionary paper, we provide an overview of the evolution of services, XaaS, agents, and the convergence of agentic AI with metaverse services. We discuss the role and key principles of agentic services in metaverse environments and introduce AMServ and Meta-AaaS as a new “as-a-Service” paradigm. We then present their characteristics, underlying principles, architectures and illustrate their potential through representative application scenarios. Finally, we outline several important research challenges and future directions for AMServ and Meta-AaaS.

\subsection{Background and Evolution of Agents}

The concept of an agent in computing dates back to the late 1950s, when John McCarthy proposed the advice taker as an early prototype of what was later called an 
agent~\cite{mccarthy1960programs}. The term itself was formalized in Hewitt’s actor model in the 1970s~\cite{hewitt1973universal}. Today, an agent is broadly understood as an entity that perceives its environment and acts autonomously on behalf of a user or system to achieve specific goals. Based on the surveys on development trends of agent techniques and self-evolving agents, the evolution process of AI agents can be classified into four stages: traditional agents, learning-based intelligent agents, LLM-empowered agents, and self-evolving agents~\cite{xi2025rise, fang2025comprehensive, gao2025survey}.

The traditional agents were primarily grounded in symbolic reasoning and multi-agent systems. Wooldridge and Jennings characterized agents through four core properties, namely autonomy, social ability, reactivity, 
and pro-activeness~\cite{wooldridge1995intelligent}, which laid the theoretical foundation for architectures such as the Belief-Desire-Intention (BDI) model~\cite{rao1995bdi}. 
% Agents in this stage relied on hand-crafted knowledge representations and operated in predefined action spaces, with intelligence largely encoded by human designers.
This type of Agents relied on hand-crafted knowledge representations and operated in predefined action spaces, with intelligence largely encoded by human designers.

The intelligent agents emerged with the rise of data-driven learning, especially deep reinforcement learning. Deep Q-Networks (DQN) demonstrated end-to-end policy learning directly from high-dimensional raw inputs~\cite{mnih2015human}, while AlphaGo combined deep neural networks with Monte Carlo Tree Search to achieve long-horizon planning over 
large state spaces~\cite{silver2016mastering}. Compared to traditional agents, this type of agents significantly improved learning and decision-making capabilities. However, such agents remained largely confined to specific task environments and closed action spaces.

The Large Language Model (LLM)-empowered agents marked a more fundamental shift by introducing large-scale pretrained LLMs into the agent loop. Chain-of-Thought prompting~\cite{wei2022chain} and ReAct~\cite{yao2022react} enabled explicit reasoning and tool invocation, while Toolformer~\cite{schick2023toolformer}, Reflexion~\cite{shinn2023reflexion}, and Generative Agents~\cite{park2023generative} further extended tool-use learning, self-reflection, 
and emergent social behavior~\cite{xi2025rise}. This type of agents substantially broadened the interaction and action space of agents, yet such agents were still limited by static model weights, fragile long-horizon planning, and insufficient persistent autonomy.

The self-evolving agents have recently emerged as a nascent paradigm that shifts intelligence from a statically configured foundation model toward an agent capable of autonomously refining its own capabilities through 
lifelong environmental interaction~\cite{fang2025comprehensive}. This paradigm emphasizes proactive goal pursuit, persistent memory, multi-agent self-organization, and autonomous optimization of prompts, memories, and tool-use strategies, with planning and reasoning progressively internalized into the agent’s policy~\cite{silver2025welcome}. 
Table~\ref{tab:agents} summarizes the evolutionary trajectory of these types of agents.

\begin{table*}[!t]
\centering
\caption{Comparison of the Four Evolutionary Types of Agents}
\label{tab:agents}
\renewcommand{\arraystretch}{1.35}
\renewcommand{\tabularxcolumn}[1]{m{#1}}
\setlength{\tabcolsep}{4pt}
\small
\begin{tabularx}{\textwidth}{
>{\centering\arraybackslash}m{0.12\textwidth}
>{\raggedright\arraybackslash}X
>{\raggedright\arraybackslash}X
>{\raggedright\arraybackslash}X
>{\raggedright\arraybackslash}X
}
\toprule
\textbf{Feature} &
\multicolumn{1}{c}{\textbf{Traditional Agents}} &
\multicolumn{1}{c}{\textbf{Intelligent Agents}} &
\multicolumn{1}{c}{\textbf{LLM-empowered Agents}} &
\multicolumn{1}{c}{\textbf{Self-evolving Agents}} \\
\midrule

Perception &
Hand-crafted symbolic percepts; passive and task-specific &
End-to-end learning from raw high-dimensional inputs &
Multimodal input via prompts and expert-model orchestration &
Active multimodal sensing refined through lifelong experience \\[0.5em]

Reasoning &
Symbolic logic-based deliberation with modal semantics &
Sub-symbolic function approximation via deep networks &
Explicit intermediate steps via chain-of-thought prompting &
Reasoning internalized into the policy and improved from experience \\[0.5em]

Planning &
Library-based means-end reasoning; short horizon &
Long-horizon lookahead via MCTS and learned network &
Recursive task decomposition with verbal self-reflection &
Self-optimizing long-horizon planning beyond external scripts \\[0.5em]

Decision-making &
Condition-action rules or expected-utility maximization &
Argmax over learned value functions or policies &
In-context language-based decisions without weight updates &
Proactive, goal-driven decisions shaped by accumulated experience \\[0.5em]

Action &
Pre-specified discrete repertoire; no external tools &
Closed action space in trained environment only &
API invocation, tool use, and code execution &
Autonomously acquired and refined tool-use strategies in open environments \\[0.5em]

Autonomy &
Weak autonomy in hand-engineered task boundaries &
High autonomy in a single trained task &
Prompt-triggered; prone to context drift and loops &
Self-initiated, persistent operation with minimal human oversight \\[0.5em]

Learning \& Adaptation &
Largely absent; learning optional in strong notions of agent &
Offline-trained and frozen, or improved through self-play &
Fixed weights; in-context learning and verbal reflection &
Lifelong self-evolution of prompts, memories, and strategies \\[0.5em]

Interaction &
Agent-agent via KQML/FIPA-ACL speech-act protocols &
Primarily agent-environment interaction in single-agent MDP settings &
Natural-language dialogue with humans; emergent social behavior in sandboxes &
Multi-agent self-organization with co-evolving roles and topologies \\

\bottomrule
\end{tabularx}
\end{table*}

\subsection{Evolution of IT-enabled Services}

The IT-enabled services deliver software functionality as discrete, network-accessible units that encapsulate business logic behind standardized interfaces, enabling loose coupling, reusability, and interoperability across heterogeneous systems \cite{Service-Manifesto}. Building upon this foundation, they have evolved through three major generations, Web services, cloud services and agentic services, in which each has increased the level of abstraction and autonomy. Web services standardized system interoperability by encapsulating business logic as stateless, reusable APIs~\cite{papazoglou2008service, WSF2014}, laying the foundation for SOA-driven inter-system integration. Introducing virtualization and elastic resource provisioning through IaaS, PaaS, and SaaS models, cloud services transformed computing infrastructure into on-demand utilities~\cite{buyya2009cloud, WeerasiriBBSR17}. Agentic services extend this progression by embedding intelligence agents directly into the service layer, thereby enabling autonomous goal pursuit. Agentic services represent a further shift in service computing: they are persistent, goal-oriented agents capable of proactively initiating actions, maintaining internal state, reasoning about their environment, 
and adapting autonomously~\cite{deng2025agentic}. LLM-driven agents can perceive context, decompose goals, invoke tools, and adapt flexible plans. This shift moves service computing from interface-driven interaction toward goal-driven execution, where the service itself can reason and act.

\subsection{Related Work on Agentic Services and Agentic Metaverse Services}

The research on agentic services has expanded rapidly in recent years and can be traced along a trajectory from service orchestration to agent autonomy. In the Web services era, Sheng et al.~\cite{sheng2014web} provided a decade-spanning survey of the service composition lifecycle, covering standards and representative prototypes, and identified emerging demands for dynamism, personalization, and reliability in service ecosystems. Xu et al.~\cite{xu2015big} subsequently introduced the notion of “Big Service”, reconceptualizing complex, cross-domain, cross-network service ecosystems as a new form of the Internet of Services, laying the conceptual groundwork for the transition from passive service invocation to proactive agent-driven collaboration.

The advent of LLMs catalyzed a more fundamental shift, elevating agents from auxiliary tools to autonomous entities that are capable of closed-loop perception, reasoning, planning, and action \cite{Aiello25}. HuggingGPT~\cite{shen2023hugginggpt} was one of the earliest systems to position an LLM as a controller for dynamically orchestrating heterogeneous model services. The platforms such as AutoGen~\cite{wu2024autogen} and CAMEL~\cite{li2023camel} provided general-purpose multi-agent platforms for dialogue orchestration and role-based cooperation, while Sapkota et al.~\cite{sapkota2025ai} clarified the distinction between monolithic AI agents and multi-agent agentic AI systems.

More recent work has moved toward a more rigorous formalization of the agentic service paradigm. Deng et al.~\cite{deng2025agentic} introduced agentic services computing, reconceiving services as LLM-driven autonomous agents and defining a four-phase lifecycle of design, deployment, execution, and evolution, organized around four research dimensions: \textit{perception}, \textit{decision-making}, \textit{collaboration}, and \textit{trustworthiness}. Zhu et al.~\cite{zhu2025agent} proposed AaaS-AN, extending the Role, Goal, Process, and Service (RGPS) meta-model~\cite{iso_iec_tr_19763-9_2015} with an Agent Network abstraction to enable networked interoperability and dynamic scheduling of agentic services.

%michael: will add the metaverse service SI 
The convergence of agentic AI and metaverse environments has begun to attract research attention. 
Xu et al.~\cite{xu2023metaverse, xu2024metaverse, Sheng-TSC-metaverse} identified the key characteristics of metaverse services, including immersive interaction, virtual-real fusion, and cross-world service aggregation, and identified the new challenges these characteristics present to service computing. 
Xiao et al.~\cite{xiao2025toward} proposed an agentic AI networking framework and demonstrated its application in metaverse and digital twin environments, in which LLM-powered agents perform autonomous and goal-oriented service delivery across physical-virtual fused spaces. However, the research into the normalized paradigm of the AMServ, such as the Meta-AaaS, remains an open problem. In this paper, we focus on the model of the AMServ and the paradigm of the Meta-AaaS.

\section{Agentic Services and AaaS in Metaverse}

This section will review the evolution process from SaaS into AaaS, introduce the concepts of agentic services and AaaS, metaverse and metaverse services, describe the roles of agents in the metaverse, and present concepts of 
%the 
AMServ and 
%the 
Meta-AaaS.

\subsection{Evolution from SaaS into AaaS}

The “as-a-service” paradigm has advanced through successive layers of abstraction, progressing from application delivery, infrastructure provisioning, to autonomous task execution. The SaaS paradigm has a deep influence on service computing. SaaS enabled users to access software over the Internet without local installation, with the provider managing underlying infrastructure across multiple tenants~\cite{10.5555/2206223}. XaaS further extended this model to cover a broad range of IT resources, including computing, storage, databases, and networking capabilities~\cite{duan2015everything}. Before agentic services, service execution was mainly initiated and directed step by step by users, with systems responding to explicit requests.

AaaS has emerged as a new stage of this evolution, introducing LLM-driven agents as the primary service unit. Users express goals in natural language or through other natural interaction ways, while agents plan, invoke tools, and dynamically adjust their actions during task execution. The service thereby shifts from executing predefined functions to autonomously pursuing desired outcomes. AaaS supports the dynamic agent composition. An agent may recruit other agents mid-task, assembling an ad hoc network to handle complexity that no pre-wired pipeline could anticipate. Table~\ref{tab:saas-aaas} shows the comparison between SaaS and AaaS.

\begin{table*}[!t]
\centering
\caption{Systematic Comparison of SaaS and AaaS}
\label{tab:saas-aaas}
\renewcommand{\arraystretch}{1.35}
\renewcommand{\tabularxcolumn}[1]{m{#1}}
\setlength{\tabcolsep}{5pt}
\small
\begin{tabularx}{\textwidth}{
>{\centering\arraybackslash}m{0.16\textwidth}
>{\raggedright\arraybackslash}X
>{\raggedright\arraybackslash}X
}
\toprule
\textbf{Dimension} &
\multicolumn{1}{c}{\textbf{SaaS}} &
\multicolumn{1}{c}{\textbf{AaaS}} \\
\midrule

Triggering mode &
Reactive; waits for explicit user request / click / API call &
Proactive and event-driven; self-triggers on goals, schedules, environment signals, or peer-agent calls \\[0.5em]

Autonomy &
None to low; executes deterministic business logic written by developers &
High; perceives environment, decomposes goals, selects actions with minimal human oversight \\[0.5em]

Context awareness &
Stateless per request or session-scoped; context is whatever the user provides &
Reasons over persistent memory, tool outputs, retrieved documents, and environmental feedback before acting \\[0.5em]

Decision / planning &
Hard-coded control flow; no runtime planning &
LLM-driven planning, reflection, tool selection, multi-step reasoning \\[0.5em]

Interaction mode &
GUI / REST API; human-in-the-loop for every step &
Natural-language goal in, outcome out; agent-to-agent messaging; human-on-the-loop supervision \\[0.5em]

Service granularity &
Whole application or feature &
Task- or role-scoped micro-capability composable into agent networks \\[0.5em]

State persistence &
Application-managed per-tenant data; no cognitive state &
Adds cognitive state: short-term working memory, long-term memory, episodic traces across sessions \\[0.5em]

Composability &
Integrations via APIs / webhooks / iPaaS; mostly static pipelines &
Agents call other agents dynamically, enabling multi-agent collaboration and marketplace-style assembly \\[0.5em]

Trust / reliability concerns &
Availability, data security, multi-tenant isolation, vendor lock-in &
All of SaaS's concerns, plus hallucination, unsafe tool use, prompt injection, non-determinism \\[0.5em]

Evaluation paradigm &
Functional tests, SLAs on uptime / latency &
Eval-driven development; objective + LLM-as-judge evaluations with error-analysis loops \\

\bottomrule
\end{tabularx}
\end{table*}

\subsection{Concepts of Agentic Services and AaaS}

An agentic service can be defined as a goal-conditioned autonomous entity that perceives its environment, reasons about tasks, and acts to pursue objectives. This definition operates at the behaviour level and describes what the entity does, including perception, reasoning, and action toward objectives. Any system that exhibits autonomy, perception-reasoning-action loops, and adaptation to environmental feedback qualifies as an agentic service, regardless of its physical form or deployment context.

AaaS (Agent-as-a-Service) denotes a service delivery mode for agentic capabilities under the XaaS paradigm. In this paradigm, agent capabilities are encapsulated as service units with explicit descriptions, deployment support, and invocation interfaces. 
% External users, applications, and domain scenarios can access
Users and applications can access these units through agentic APIs or multimodal interfaces under provider-consumer agreements. A service entity may expose perception, cognition, decision-making, execution, collaboration, or content generation capabilities. Multiple agentic services can also be composed into a large composite agentic service. AaaS therefore applies established mechanisms from services computing, including service registration, discovery, orchestration, composition, Service Level Agreements (SLAs), monitoring, and value management, to autonomous agents.

\begin{figure}[!htbp]
\centering
\includegraphics[width=\columnwidth]{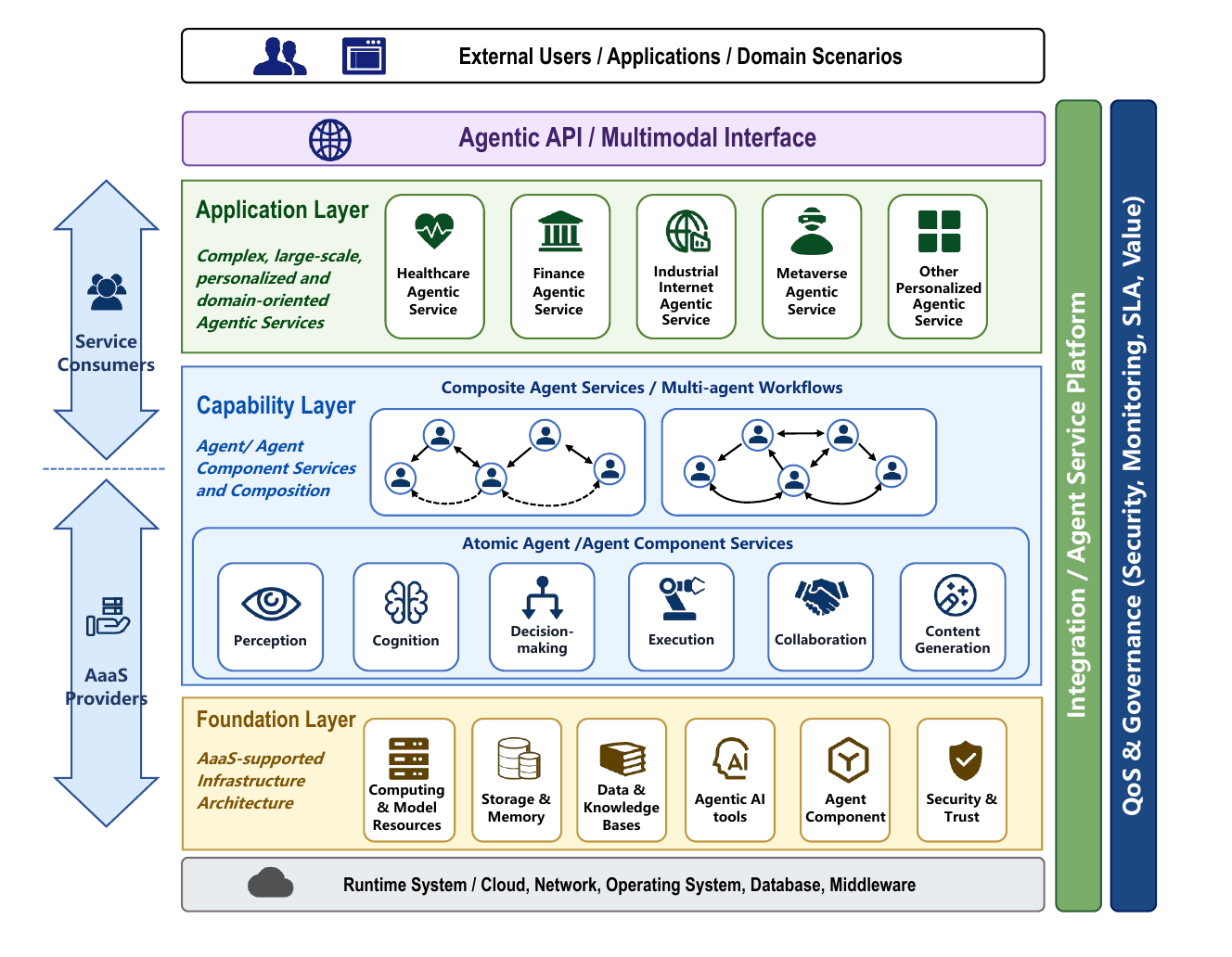}
\vspace{-7mm}
\caption{The conceptual architecture of AaaS}
\label{fig1}
\vspace{-5mm}
\end{figure}

The architecture of AaaS consists of three functional layers built on a runtime system, with cross-layer platform and governance support, as shown in Fig.~\ref{fig1}. The runtime system provides the general execution substrate, including cloud infrastructure, networks, operating systems, databases, and middleware. Based on this substrate, the foundation layer provides the AaaS-supported infrastructure architecture, including computing and model resources, storage and memory, data and knowledge bases, agentic AI tools, agent components, and security and trust mechanisms. These elements support the deployment, execution, and operation of agentic services. 

% The capability layer organizes agent abilities as atomic agent services or agent-component services. 
The capability layer organizes agent abilities as atomic agent services or component agents services.
Typical atomic agentic services include perception, cognition, decision-making, execution, collaboration, and content generation. These services can be invoked individually or composed through orchestration mechanisms into composite agentic services. This layer provides a service-level abstraction for agent capabilities, making them reusable, interoperable, and composable across different application contexts. 

The application layer assembles these capabilities into complex, large-scale, personalized, and domain-oriented Agentic Services. Representative examples include healthcare agentic services, finance agentic services, industrial agentic services, Agentic Metaverse Services, and other personalized agentic services. 
% These services are exposed to external users, applications, and domain scenarios through agentic APIs and multimodal interfaces, which support interaction through text, image, voice, gesture, and other channels. 
These services are exposed to users and applications through agentic APIs and multimodal interfaces, which support interaction through text, image, voice, gesture, and other channels.

In addition to the multi-layer architecture, AaaS relies on two types of cross-layer support. The integration and agentic service platform provides mechanisms for service development, deployment, composition, integration, operation, and lifecycle management. The QoS and governance environment supports security, monitoring, SLA management, trust assurance, and value realization. These cross-layer mechanisms connect service consumers with AaaS providers and provide the operational basis for quality control, accountability, and service management.

A componentized architecture supports on-demand invocation and independent scaling of capability units. A pay-as-you-go or value-based service model can also reduce the cost of adopting agent technologies for organizations that do not maintain full agent stacks internally. Multimodal interfaces allow users to interact with AaaS through natural channels, while standardized agentic APIs support coordination among AaaS instances and external systems. After deployment, learning mechanisms can update decision policies or service strategies under governance constraints. AaaS may support proactive service delivery, where multimodal perception observes user context, generative AI-based inference identifies potential needs, and orchestration mechanisms select and invoke appropriate services with reduced explicit user requests. This AaaS architecture provides a basis for moving from user-initiated service access toward context-aware service delivery.

\subsection{Metaverse and Metaverse Services}

The concept of metaverse was first introduced by Neal Stephenson in his 1992 science fiction novel \textit{Snow Crash}, which depicted a three-dimensional (3D) virtual space in which humans interact through programmable avatars~\cite{stephenson1994snow}. With the rapid advancement of next-generation digital technologies, including AI, blockchain, cloud computing, edge computing, 5G/6G communications, VR/AR/MR, and the Internet of Things, the metaverse has progressively evolved from a speculative fiction concept into a technically feasible ecosystem. While no consensus definition exists in the literature, the metaverse is widely understood as a virtual social ecosystem constructed through digital technologies that interacts deeply with the physical world, encompassing diverse existence modalities such as digital twins, digital natives, and cyber-physical symbiosis. The metaverse can be understood as a complex socio-technical platform characterized by real-time rendering, interoperability, persistent state, and massively concurrent 
user access~\cite{ball2022metaverse, Metaverse-CSUR2013, Wang-metaverse23}.

\begin{figure*}[!t]
\centering
\includegraphics[width=0.95\textwidth]{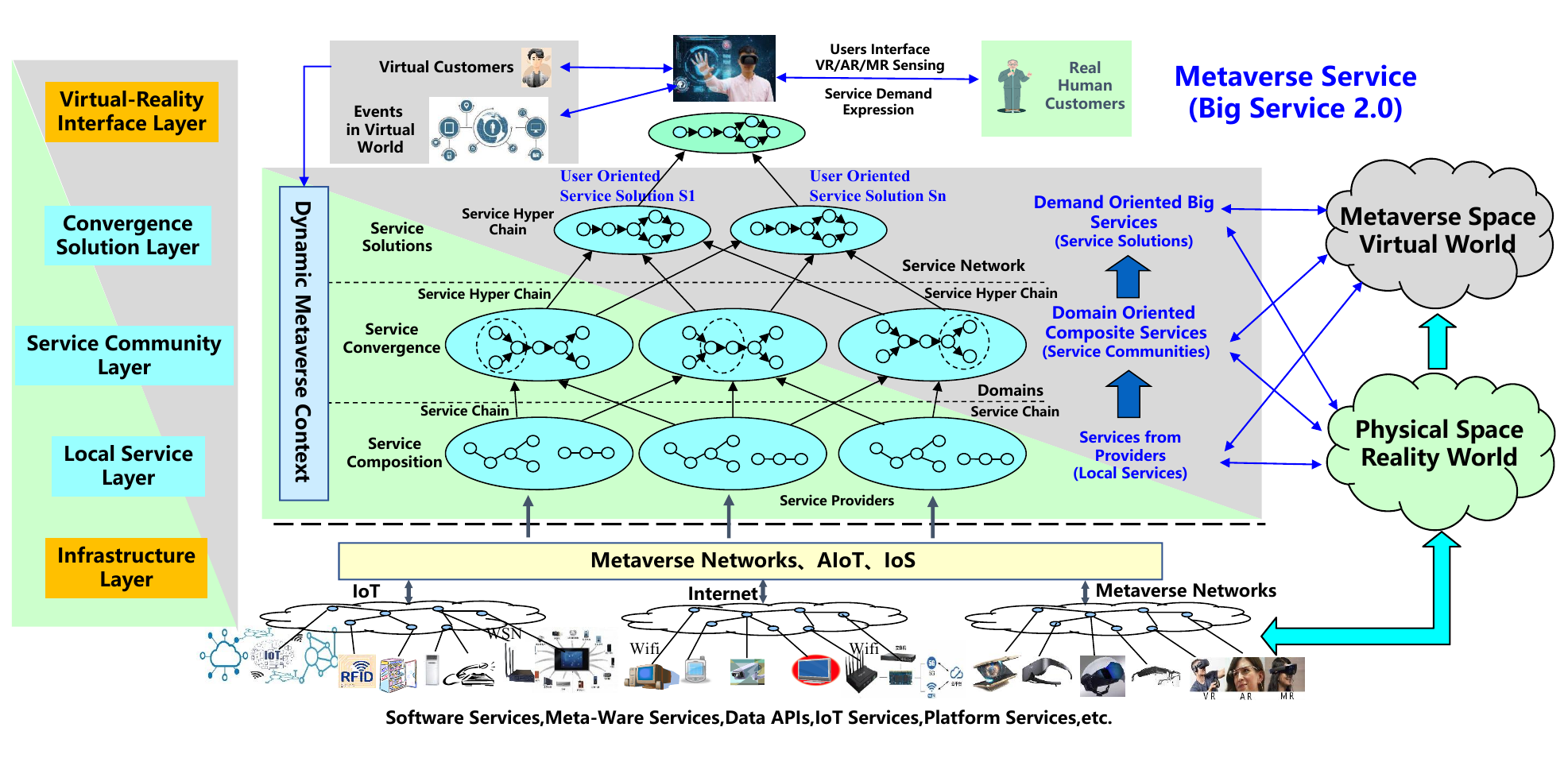}
\vspace{-5mm}
\caption{The conceptual architecture of the Metaverse services}
\label{fig2}
\vspace{-5mm}
\end{figure*}

In the metaverse ecosystem, social activities and business processes are organized and delivered in services, giving rise to the notion of Metaverse services. A Metaverse service can be defined as a composite service paradigm composed of large-scale, heterogeneous digital services operating across both virtual and cyber-physical spaces, designed to fulfill the service demands of digital avatars, digital twins, and digital natives through cross-network, cross-domain, cross-regional, and cross-world aggregation and collaboration~\cite{xu2023metaverse, xu2024metaverse}. Different from the conventional Internet services, the Metaverse service entities expand beyond software services to encompass digital twins and Metaverse-Ware~\cite{xu2023metaverse, xu2024metaverse}. The service experience shifts from traditional on-demand interaction to immersive, mixed-reality engagement. Accordingly, Metaverse services can be regarded as an upgraded instantiation of the Big Service paradigm \cite{xu2015big}, i.e., Big Service 2.0, characterized by greater diversity in service modalities, increased complexity in business content, and new challenges and open research problems for the fields of service computing and service engineering~\cite{xu2023metaverse, xu2024metaverse, ShiXXW22}. Fig.~\ref{fig2} illustrates the conceptual architecture of the Metaverse services.

From a hierarchical service-organization perspective, the metaverse service system consists of five layers: infrastructure, local services, service communities, convergence service solutions, and virtual-reality interfaces. In the infrastructure layer, there are basic resources, including software, virtual environment, IoT, cloud, data, communication, and platform services \cite{BouguettayaSBNM21}. The local service layer integrates these resources into composite services for metaverse business applications, while the service community layer aggregates them into cross-world communities to support virtual-physical integration. The convergence solution layer further combines services across networks, domains, and worlds to address diverse user requirements~\cite{benatallah2003self}. Finally, the virtual-reality interface layer delivers personalized, large-scale metaverse services by adapting integrated service elements to user needs and experiences~\cite{xu2023metaverse, xu2024metaverse}.

\subsection{The Roles of Agents in the Metaverse}

In the Metaverse service ecosystem, agents serve as the functional entities bridging users, virtual environments, the physical world, and service systems. Arunkumar et al.~\cite{v2026agenticartificialintelligenceai} proposed a unified taxonomy of agentic AI systems covering perception, planning, action, tool use, and collaboration. Based on this, we propose in this paper a role-based taxonomy classifying metaverse agents into six types: \textit{perceptual agents}, \textit{planning and decision-making agents}, \textit{execution agents}, \textit{content generation agents}, \textit{social collaboration agents}, and \textit{evaluation and governance agents}, where the latter two are introduced to address requirements specific to the metaverse service ecosystem. Regarding the functional responsibilities of each role, the types of metaverse agents are elaborated as follows.

%\paragraph{
\vspace{1mm}
\noindent \textbf{Perceptual Agents}. Perceptual agents are responsible for the real-time acquisition and interpretation of multimodal inputs within the metaverse, including user speech, text, gestures, interface interactions, scene states, and cyber-physical synchronization data, providing the basic input for downstream reasoning and service adaptation.

%\paragraph{
\vspace{1mm}
\noindent \textbf{Planning and Decision-Making Agents}. Planning and decision-making agents operate on perceptual outputs to perform task decomposition, resource coordination, decision-making, and strategy synthesis, translating user intent and contextual constraints into actionable service plans. Together with perceptual agents, they constitute the cognitive foundation of metaverse services, supporting the transition from reactive response to proactive adaptation.

%\paragraph{
\vspace{1mm}
\noindent \textbf{Execution Agents}. Execution agents realize service solutions by driving virtual operations, invoking APIs, and orchestrating workflows, and may additionally enable real-world actuation through digital twins, edge devices, or intelligent terminals.

%\paragraph{
\vspace{1mm}
\noindent \textbf{Content Generation Agents}. Content generation agents produce the diverse digital assets required by metaverse services, including textual descriptions, spoken dialogue, images, videos, three-dimensional assets, and narrative storyboards, supplying the content foundation for immersive and personalized service delivery.

%\paragraph{
\vspace{1mm}
\noindent \textbf{Social Collaboration Agents}. Social collaboration agents support negotiation, task allocation, and cooperative execution among multiple agents and between agents and human users.

%\paragraph{
\vspace{1mm}
\noindent \textbf{Evaluation and Governance Agents}. Evaluation and governance agents are responsible for safety auditing, quality assessment, behavioral constraint enforcement, and rule execution, ensuring the coordinated, orderly, and compliant operation of the overall system.

The above types of agents denote basic functional roles that can be combined within concrete metaverse agents or agentic services. In practical metaverse service scenarios, agentic services usually operate in a composite form, dynamically combining multiple roles to achieve specific service goals. 
For example, a healthcare agentic service in the metaverse may simultaneously perceive user behaviour and physiological signals, generate consultation content, coordinate medical resources, execute service workflows, and comply with safety and privacy constraints. This goal-driven composition explains why a concrete agentic service often integrates multiple basic roles to complete specific service goals \cite{LemosDB16}.

From an architectural perspective, these basic roles constitute a service chain from perception and cognition to execution, content generation, collaboration, and governance. As composable functional building blocks, they provide a role-level abstraction for characterizing the functional composition of special-purpose metaverse agents and supporting their design, composition, and orchestration. From this perspective, the proposed taxonomy offers a basic structure for constructing and continuously evolving composite agents in the metaverse service ecosystem.

\subsection{Concepts of Agentic Metaverse Services and the Meta-AaaS}

% Agentic Metaverse services can be defined as a class of intelligent, agent-driven services that operate within metaverse environments to support proactive, scene-aware, and immersive service delivery across virtual spaces and cyber-physical hybrid domains. In essence, Agentic Metaverse services represent the systematic realization of multi-role agent collaboration within the continuously evolving and generative context of the metaverse. Agentic Metaverse services thus operate at the behavioral level, characterizing what agent-driven services do within the metaverse, while the Meta-AaaS addresses how such services are packaged, discovered, and delivered under a provider-consumer contract. Regarding the paradigm through which these services are organized and delivered, the Meta-AaaS constitutes a dedicated implementation framework.

% The Meta-AaaS can be defined as an emerging service paradigm that encapsulates agent capabilities into discrete, reusable, and composable service units, deployable and invocable on demand within metaverse environments. Building on the AaaS foundation established above, the Meta-AaaS situates agent service units within the metaverse, where service logic must additionally accommodate spatial dynamics, identity continuity, and embodied interaction. In essence, the Meta-AaaS is an agent-native service architecture where the metaverse functions simultaneously as the operational space and as a defining constraint on service design. Fig.~\ref{fig3} illustrates the conceptual architecture of the Meta-AaaS.

\begin{figure*}[!tb]
\centering
\includegraphics[width=\textwidth]{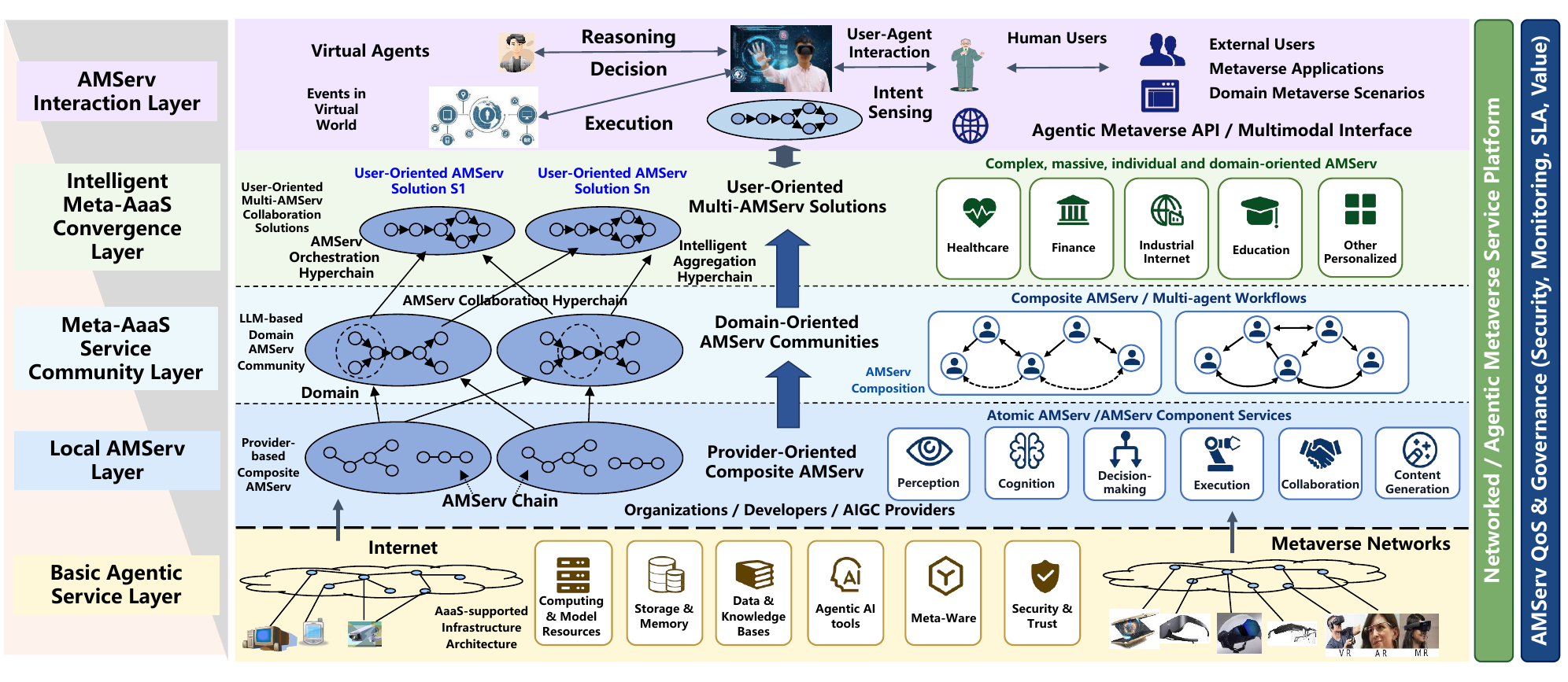}
\vspace{-9mm}
\caption{The conceptual architecture of the Meta-AaaS}
\label{fig3}
\vspace{-5mm}
\end{figure*}

Agentic Metaverse Services (AMServ) can be defined as the intelligent, agent-driven complicated services operating in metaverse environments, to support autonomous, proactive, scene-aware, and immersive service delivery across virtual spaces and cyber-physical hybrid domains. They proactively perceive metaverse states and events, interpret user’s intentions, set the service goals, converge and coordinate virtual and physical resources, and act and deliver services through embodied or multimodal interfaces. AMServ describe the function of agent-driven services in the metaverse, while the Meta-AaaS provides the as-a-Service architecture for their development, encapsulation, deployment, invocation, composition, execution, and governance. Capabilities such as perception, interaction, planning, execution, content generation, collaboration, and governance can be combined to form the composite AMServ. For example, a multimodal interaction capability can support different applications, e.g., virtual tourism guidance, classroom tutoring, or healthcare consultation, 
%etc., 
when combined with different domain knowledge, tools, and interfaces. The AMServ can form the service forms discussed as follows.

%\paragraph{
\vspace{1mm}
\noindent \textbf{General Agentic Metaverse Services}. This form provides reusable frames and capabilities of AMServ, including natural language interaction, multimodal interface adaptation, task planning, tool use, execution control, and basic service coordination. This type of 
%the 
AMServ function as general agent service frames 
%that 
can be adapted to different metaverse scenarios.

%\paragraph{
\vspace{1mm}
\noindent \textbf{Domain-oriented Agentic Metaverse Services}. This form integrates agent capabilities with domain knowledge, business rules, and metaverse-specific elements such as people, events, objects, and scenes for specific domains. They provide services adapted to the application scenarios such as education, healthcare, eldercare, cultural tourism, industrial operations, and urban governance. For example, in a metaverse education ecosystem, the coordinated interaction among teacher agents, student agents, and NPCs represents a typical educational domain-oriented service.

%\paragraph{
\vspace{1mm}
\noindent \textbf{Collaborative Agentic Metaverse Services}. This form emphasizes multi-agent coordination across domains, platforms, and virtual-physical spaces. Multiple agents may divide roles, exchange information, coordinate actions, and jointly complete service tasks across interconnected metaverse environments. The collaborative services can be realized by means of the Meta-AaaS framework and operational platforms.

AMServ has the following key characteristics:

\begin{itemize}
    \item \textbf{Autonomy:} AMServ can independently perceive environmental states and events, interpret user’s intentions, set the service goals, and perform actions with minimal or no human intervention.
    \item \textbf{Proactivity:} AMServ can proactively identify latent user needs from contextual cues, such as user location, behavior trajectories, scene changes, and spatial events, and provide the adaptive operation actions timely.
    \item \textbf{Embodiment:} AMServ can appear and act through avatars, NPCs, virtual humans, robots, or digital twins, enabling them to participate in the 3D virtual spaces and cyber-physical environments, providing users with embodied experiences.
    \item \textbf{Cross-time-space convergence:} The service components can be converged and coordinated across time, virtual spaces, physical spaces, platforms, domains, and worlds, to constitute complex AMServ.
    \item \textbf{Collaboration:} The multi-agents can collaborate and coordinate with each other in shared or interconnected service processes for a certain task goal in metaverse ecosystem.
    \item \textbf{Evolvability:} AMServ can actively evolve themselves to adjust their strategies, interaction patterns, and domain adaptation through accumulated data, user feedback, and scene experience.
\end{itemize}

Meta-AaaS, as an “as-a-Service” paradigm, can be defined as an implementation architecture and operation environment for  AMServ. Meta-AaaS encapsulates agent capabilities into configurable, invocable, reusable service entities, composes the basic agentic services into the complicated AMServ through “as-a-Service” paradigm, and supports their operation in the metaverse environment. The typical functions of Meta-AaaS include dynamic agentic service configuration in virtual space, digital agent identity continuity, convergence and composition of AMServ, AMServ construction, multi-agent coordination, AMServ operation, embodied agentic service interaction and multimodal interfaces, and cyber-physical coordination and collaboration of the AMServ in metaverse ecosystems, etc. 

Meta-AaaS includes both a development and construction environment and a runtime and maintenance environment, as shown in Fig.~\ref{fig3}. The former supports service requirement identification, agent capability modelling, agentic service encapsulation, agent role configuration, agentic service composition, agent-as-a-service definition, knowledge and tool binding, multimodal interface specification, agentic service testing, while the latter supports service registration, discovery, invocation, operation, scheduling, coordination, monitoring, maintenance, evolution, feedback collection, and governance. Compared to the AMServ which describe the function and behavior of agentic services in metaverse environments, the Meta-AaaS provides the mechanism of architecture, implementation and operation in which the agentic services are constructed, deployed, composed, managed, and governed.

\section{Principles and Applications of Agentic Metaverse Services and Meta-AaaS}

\subsection{Principles of Agentic Metaverse Services and Meta-AaaS}

\begin{figure}[!tbp]
\centering
\includegraphics[width=\columnwidth]{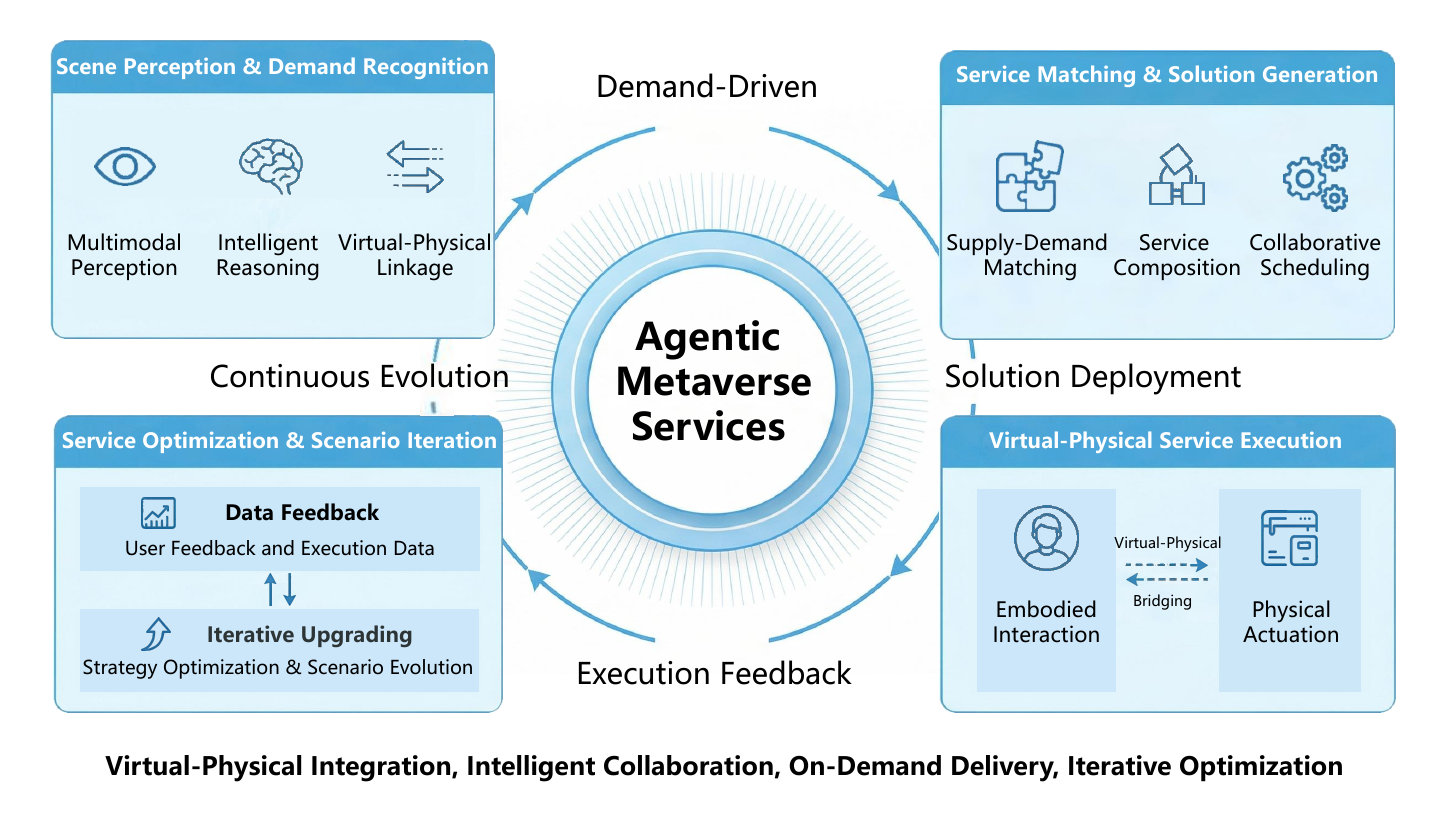}
\vspace{-5mm}
\caption{The principle of AMServ}
\label{fig4}
\vspace{-5mm}
\end{figure}

AMServ operates through four connected stages. At the first stage, \textit{Scene Perception and Demand Recognition}, the AMServ system captures multi-modal signals from virtual and physical spaces, such as scene states, user behaviors, digital twin data, and device status, senses and recognizes the service demands related to specific users, avatars, tasks, or scenarios. In the second stage, \textit{Service Matching and Solution Generation}, the system matches the demands and service supply, and plans the adaptive service solutions by using available agent capabilities, data resources, and digital assets. The recognized demands are then organized into executable service plans for domain-specific metaverse scenarios. In the third stage, \textit{Virtual-Physical Service Execution}, the service plans are deployed through Metaverse-Ware and delivered through avatars, NPCs, digital humans, and digital twins. When required, physical actuation is triggered through connected devices and terminals. Finally, at the fourth stage, \textit{Service Optimization and Scenario Iteration}, execution feedback, including interaction traces, service logs, and digital twin states, is analyzed to refine matching strategies, composition patterns, and planning policies. This stage supports the evolved improvement of service behavior and metaverse scenarios. This operational principle of AMServ is illustrated in Fig.~\ref{fig4}.

\begin{figure}[!htbp]
\vspace{-2mm}
\centering
\includegraphics[width=\columnwidth]{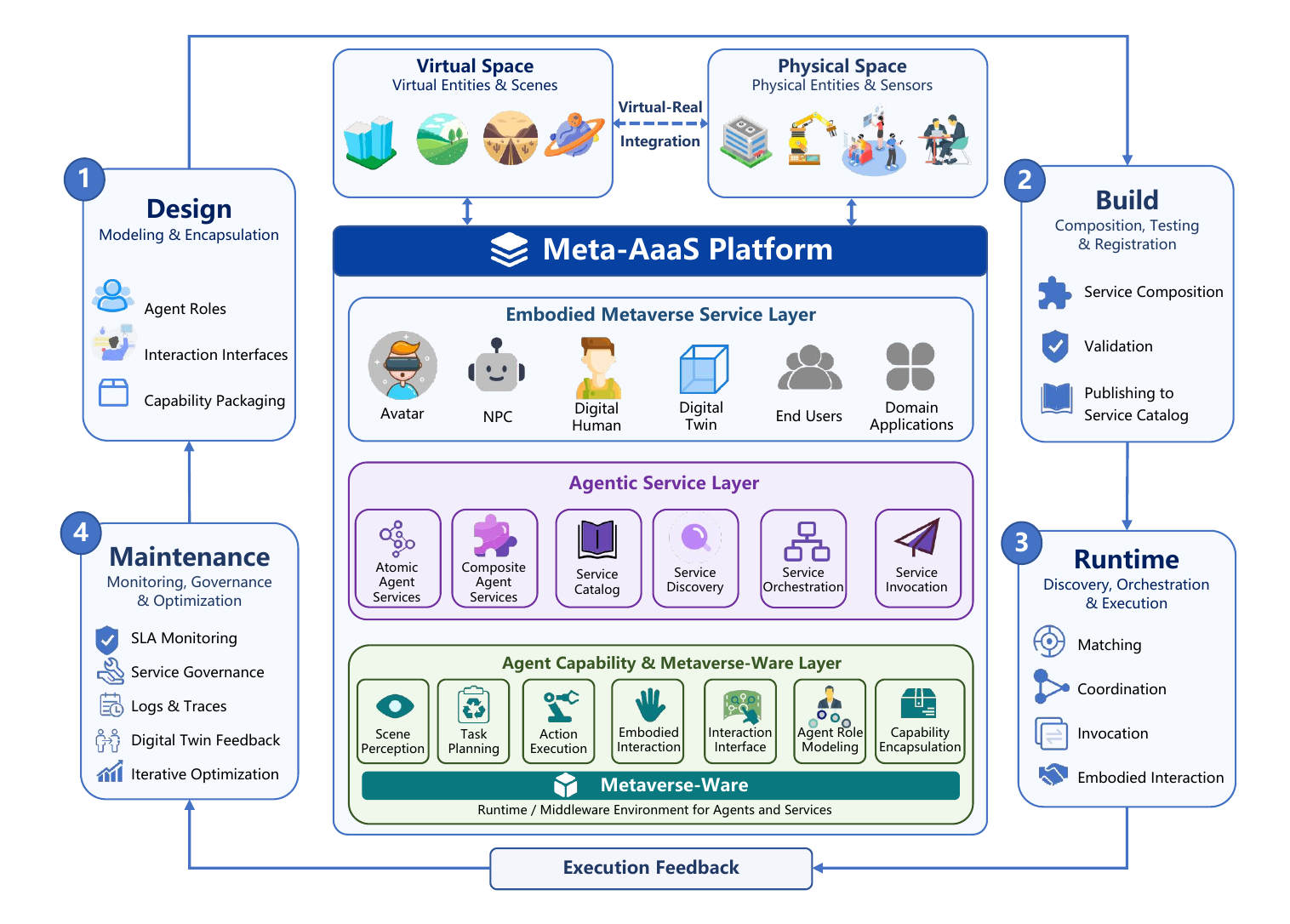}
\vspace{-6mm}
\caption{The architecture of Meta-AaaS}
\label{fig5}
\vspace{-3mm}
\end{figure}

% To realize AMServ, Meta-AaaS operates as a platform that supports service design, construction, operation, maintenance, and evolution, as four phases as shown in Fig.
To realize AMServ, Meta-AaaS operates as a platform that supports four phases: service design, construction, operation, and maintenance, while enabling continuous evolution throughout the lifecycle, as shown in Fig.~\ref{fig5}. The first phase is the \textit{Design Phase}, which covers agent service modelling and capability encapsulation. Developers define agent roles, specify interaction interfaces, and package agent capabilities, such as scene perception, task planning, action execution, and embodied interaction, as service entities aligned with the three-layer architecture of Meta-AaaS. The second phase is the \textit{Build Phase}, which covers service composition, testing, and registration. In this phase, atomic and composite agentic services are composed and constructed, validated, and published to the Meta-AaaS service catalogue for later discovery and invocation in metaverse application scenarios. The third phase is the \textit{Runtime Phase}, which handles service discovery, invocation, and execution. Deployed agentic services are matched and coordinated to provide domain-oriented agentic services to end users through embodied metaverse interfaces, such as avatars, NPCs, digital humans, and digital twins. Finally, the \textit{Maintenance Phase} manages SLA monitoring, service governance, and iterative optimization based on execution feedback, including interaction traces, service logs, and digital twin states. These activities support reliable service operation and adaptation to changing metaverse scenes. The four phases provide the platform basis for organizing the AMServ into modular, manageable, and evolvable service delivery.

\subsection{Applications of Agentic Metaverse Services and Meta-AaaS}

Among numerous applications of AMServ and Meta-AaaS, five typical examples are briefly mentioned here to illustrate how Meta-AaaS 
%realizes 
can be applied to realize 
AMServ in practice.

%\paragraph{
\vspace{1mm}
\noindent \textbf{Healthcare and Elderly-Care}. In the healthcare and elderly-care domains, AMServ can support continuous monitoring of users’ physiological states and daily behaviors, enabling proactive intervention before explicit help requests are made. Such agents may assist with medication reminders, anomaly detection, emergency alerts, and coordination with family members or clinical systems, thereby improving the timeliness and personalization of care services~\cite{sheng_2026_aging}. Moreover, in immersive metaverse environments, agents can embody personalized avatars to deliver virtual companionship and VR-based rehabilitation, while leveraging users’ digital twins to simulate health trajectories and enable predictive interventions beyond conventional telemedicine.

%\paragraph{
\vspace{1mm}
\noindent \textbf{Industrial Manufacturing}. In the field of industrial manufacturing, AMServ can enable continuous inspection and intelligent supervision of production processes through digital twins, sensor networks, and autonomous maintenance agents. For instance, in a digital twin-driven manufacturing system, agents leverage the virtual replicas to simulate potential operating scenarios and pre-validate maintenance strategies. The optimized strategies are then deployed to physical equipment, enabling risk-free decision-making and zero-disruption optimization on the production line. By identifying abnormal states early and generating timely maintenance recommendations, the agentic service systems help improve equipment reliability and reduce downtime \cite{IndustrialMeta}.

%\paragraph{
\vspace{1mm}
\noindent \textbf{Education}. In the education domain,  AMServ can provide agent services for adaptive learning based on students’ real-time performance, interaction history, and scene-related needs. Rather than delivering uniform content to all learners, the agents can dynamically adjust pace, difficulty, explanation style, and interaction modes according to individual progress, thereby supporting more personalized and learner-centered learning experiences. In immersive metaverse classrooms, agents can act as intelligent NPC tutors and peer collaborators. By contextualizing abstract concepts through interactive 3D simulations and virtual laboratories, they enable experiential and situated learning that is difficult to achieve in conventional 
online education~\cite{xu2024metaverse}.

%\paragraph{
\vspace{1mm}
\noindent \textbf{Entertainment}. In the field of entertainment, AMServ can support the dynamic generation of immersive experiences by creating scenes, narratives, characters, and interactions in response to users’ behaviors and preferences. In interactive gaming, agents can serve as intelligent non-player characters that perform with distinct personalities, perceive players’ emotions, and improvise personalized dialogues in real time. Beyond individual roles, multiple agents can collaboratively orchestrate role-playing, procedurally generate scenes and quests, and evolve branching storylines while maintaining a coherent worldview. Users may participate not only as consumers of pre-authored content, but also as active participants in interactive and co-created experiences~\cite{nisiotis2025emerging}.

%\paragraph{
\vspace{1mm}
\noindent \textbf{Culture and Tourism}. In culture and tourism, AMServ can provide immersive and individualized virtual travel services for heritage exploration, destination guidance, and cultural interpretation by means of agents~\cite{catarci2024ethics, koo2023metaverse}. Through combining virtual environment rendering, multilingual interaction, and scene recommendation, such agents may improve access to geographically remote or physically inaccessible destinations and extend tourism experiences beyond traditional on-site visits. Beyond navigation and recommendation, these agents can act as virtual tour guides and cultural storytellers that reconstruct heritage scenes across time and space, inviting tourists to role-play historical figures, interact with cultural artifacts, and collectively shape narrative storylines, thereby transforming passive sightseeing into participatory co-creation of cultural experiences.

\section{New Research Topics for Agentic Metaverse Services and Meta-AaaS}

% Being in the cross field of multiple new technologies, including GenAI, LLMs, agentic techniques, metaverse, and service computing, Agentic Metaverse services represent the new paradigm of agentic services in metaverse environments. The Meta-AaaS serves as the primary delivery mechanism of this paradigm, transforming various agent capabilities into composable and deliverable services that can meet the demands of metaverse environments, including virtual-physical fusion, immersive interaction, mass individualized services, and decentralized operation. The important research topics of Agentic Metaverse services and the Meta-AaaS are discussed as follows.
Being in the 
%cross field 
intersection 
of multiple new technologies, including GenAI, LLMs, agentic techniques, metaverse, and service computing, AMServ represents the new forms of agentic services in metaverse environments. Meta-AaaS is an approach to implement AMServ as the as-a-service paradigm, transforming various agent capabilities into composable and deliverable services in metaverse environments. Several important research topics of AMServ and Meta-AaaS are discussed as follows.

%\paragraph{
\vspace{1mm}
% \noindent \textbf{Service Representation Models}. A fundamental issue is to represent agent services and their capabilities in Agentic Metaverse service environments. Existing service description approaches were not designed for agents with embodied forms, evolving behaviors, and multimodal interaction capabilities. New representation models of agentic services in metaverse should express not only the service functional capabilities, but also the features of agentic services which can be realized in the Agentic Metaverse service environment, e.g., digital avatars, virtual business, 3D scene and meta-ware made things, and virtual-physical fusion. A critical challenge is cross-world portability: how agent capabilities, reputation, and service commitments translate across heterogeneous virtual world ontologies.
\noindent \textbf{Service Representation Models}. A fundamental issue is to represent agentic services and their capabilities in the AMServ environments. In order to present agentic services with embodied forms, proactive behaviors, and multimodal interaction capabilities in the metaverse environment, the new representation models of the AMServ should express not only the service functional capabilities, but also the features of agentic services, which can be realized in the Meta-AaaS environment, e.g., digital avatars, digital natives, virtual business, 3D scene and metaverse things, and virtual-physical fusion. A critical challenge is the cross-world portability and interoperability: how can the agentic service capabilities, collaborative behaviors, service commitments, and service value translate across heterogeneous virtual worlds in the metaverse.

%\paragraph{
\vspace{1mm}
% \noindent \textbf{Active Requirement Sensing}. Unlike traditional services that wait for explicit invocation requests, Agentic Metaverse services are expected to discover and identify user demands proactively. In metaverse service environments, such demands may emerge from real or virtual user’s behavior, scene’s changes, and environmental signals, rather than only direct requests by customers. The facilities and mechanisms of proactive requirement sensing should be reasonably designed, in order to sense, infer, and predict service demands from multimodal interaction data and historical context. The persistent spatial immersion of metaverse enables longitudinal demand modeling from users’ spatio-temporal behavioral trajectories across sessions.
\noindent \textbf{Active Requirement Sensing}. Unlike traditional services that wait for explicit invocation requests, AMServ is expected to discover and identify user demands proactively. In metaverse service environments, such demands may emerge from real or virtual user’s behavior, scene’s changes, virtual trigger events, and environmental signals, rather than only direct requests by customers. The facilities and mechanisms of proactive requirement sensing by AMServ should be reasonably designed, in order to sense, infer, and predict service demands from multimodal interaction data, signals, events, and historical context. A key challenge is the multi-dimensional facility modeling for AMServ to proactively sense, discover, judge and react the encountered and the predicted requirements of the service recipients in the 3D metaverse environment.

%\paragraph{
\vspace{1mm}
% \noindent \textbf{Agent-Based Service Ecosystems}. Agentic Metaverse services will drive the facilities of metaverse service systems to transform into the paradigm of agentic services. As the collaborative feature of Agentic Metaverse services, it is important to research on multi-agents’ collaboration for agents to discover one another, establish collaboration, and evolve in an open metaverse service ecosystem. This deals with the roles of providers, developers, end users, and supporting infrastructures that enable dynamic coordination and sustainable development of metaverse service ecosystem. The novelty here is the emergence of “service ecologies” where agents may form symbiotic relationships and co-evolve capabilities beyond static supply-demand market structures.
\noindent \textbf{Agent-Based Service Ecosystems}. AMServ will drive the facilities of metaverse service systems to transform into the paradigm of agentic services. The novelty here is the emergence of “agentic service ecosystems” where agents may form symbiotic relationships and co-evolve capabilities in the metaverse service environment. Agentic services deal with the roles of providers, developers, end users, and supporting infrastructures that enable dynamic coordination and sustainable development of metaverse service ecosystems. Moreover, it is important to conduct research on multi-agent collaboration for agents to discover one another, to establish collaborations and evolve in an open metaverse service ecosystem.

%\paragraph{
\vspace{1mm}
% \noindent \textbf{Cross-Time-Space Service Convergence}. Agentic Metaverse Services often span virtual and physical spaces across different locations and temporal scenes. It is necessary to research on cross-time-space service convergence and composition dynamically under constraints such as latency, resource availability, and scene transforming in the metaverse environment. The effective composition mechanisms and efficient algorithms will be essential for ensuring continuity and adaptability in complicated cross-space service scenes in metaverse. The particular challenge lies in temporal asynchrony when virtual worlds run at divergent clock speeds relative to physical time.
\noindent \textbf{Cross-Time-Space Service Convergence}.  AMServ often spans virtual and physical spaces across different locations and temporal scenes. It is necessary to conduct research on cross-time-space dynamic agentic service convergence and composition in Meta-AaaS, under the constraints of latency, resource availability, and scene transformation in the metaverse environment. The effective composition mechanisms and efficient algorithms of AMServ will be essential for ensuring continuity and adaptability in complicated cross-space service scenes in metaverse. The interoperability and fusion of heterogeneous agentic services with asynchronous temporal characteristics from different spaces of the virtual metaverse world and physical 
%world would be 
remains 
a difficult problem.

%\paragraph{
\vspace{1mm}
% \noindent \textbf{Coordinative Service Governance}. It is also important to coordinatively manage and govern the agentic services in Agentic Metaverse service ecosystems. When multiple agents collaborate on service tasks, conflicts may arise with respect to resources, priorities, and decision outcomes. It needs a set of governance mechanisms for conflict detection, responsibility allocation, coordination control, and stable service execution in dynamic metaverse environments. Unique to the metaverse is “spatial governance”—regulating not just what agents do, but where and when they may do it in the shared 3D space.
\noindent \textbf{Coordinative Service Governance}. It is also important to collaboratively manage and govern the agentic services in Meta-AaaS ecosystems. When multiple agents collaborate on service tasks, conflicts may arise with respect to resources, priorities, and decision outcomes. It requires a set of governance mechanisms in Meta-AaaS platforms, for conflict detection, responsibility allocation, coordination control, and stable service execution in dynamic metaverse environment. Unique to the metaverse is “spatial governance”, regulating not just what agents do, but where and when they may do it in the open 3D metaverse space.

%\paragraph{
\vspace{1mm}
% \noindent \textbf{Safe and Reliable Service Operation}. Long-running autonomous agents introduce new risks for service operation, especially when they interact across multiple scenes or invoke external tools continuously. The research on safe and reliable service operation is needed for behavior auditing, anomaly detection, runtime intervention, and reliability assurance to support safe and dependable operation of Agentic Metaverse service systems. Risks are amplified by social density: a malfunctioning agent may disrupt immersive experience or trigger cascading reputation damage across agent networks.
\noindent \textbf{Safe and Reliable Service Operation}. Long-running autonomous agents introduce new risks for service operation, especially when they interact across multiple scenes or invoke external tools continuously. The research on safe and reliable service operation is necessary for behavior auditing, anomaly detection, runtime intervention, and reliability assurance to support safe and dependable operation of Meta-AaaS systems. Risks can be amplified by social density. A malfunctioning agent may disrupt immersive experience or trigger cascading reputation damage across Meta-AaaS networks.

%\paragraph{
\vspace{1mm}
% \noindent \textbf{Embodied Interaction and Perception}. Embodied interaction and perception are the distinctive features of Agentic Metaverse services, which interact in metaverse environments through avatars, digital twins, virtual entities, and agents, rather than only through text or conventional APIs. Supporting natural, multimodal, and embodied interaction based on agents or agentic services, including voice, gesture, expression, and spatial behavior, while maintaining service consistency and perceptual accuracy, remains a critical technical challenge for agentic services. The aesthetic and social appropriateness of an agent’s avatar form directly impacts service trust in ways that API latency never could.
\noindent \textbf{Embodied Interaction and Perception}. Embodied interaction and perception are the distinctive features of AMServ, which interact in metaverse environments through avatars, digital twins, virtual entities, and agents. Supporting natural, multimodal, and embodied interaction based on agents or agentic services, including voice, gesture, expression, and spatial behavior, while maintaining service consistency and perceptual accuracy, remains a critical challenge for agentic services. The aesthetic and social appropriateness of an agent’s avatar form may directly impact service trustworthiness.

%\paragraph{
\vspace{1mm}
% \noindent \textbf{Active Immersive Service Delivery}. Immersive experience of service delivery is a key metric of customer satisfaction in Agentic Metaverse services. Immersive service delivery requires not only presenting them in three-dimensional virtual environments, but also involves adapting content, pacing, and interaction styles to users’ real-time states, usage habits, and surrounding scenes in a proactive manner. The virtual world itself becomes a dynamically configurable service delivery channel where environmental ambience may be agentically adjusted. How to achieve such active and immersive service delivery by means of suitable agentic service facilities and tools remains an important research topic for Agentic Metaverse services. It is critical to preserve user autonomy when their perceptual environment is continuously agentically curated.
\noindent \textbf{Active Immersive Service Delivery}. Immersive experience of service delivery is a key metric of customer satisfaction on AMServ. Immersive delivery of AMServ requires not only presenting them in three-dimensional virtual environments, but also adapting content, pacing, and interaction styles to users’ real-time states, usage habits, and surrounding scenes in a proactive manner. The virtual world itself becomes a dynamically configurable service delivery channel where environmental ambience may be adjusted by the agentic services. How to achieve such active and immersive service delivery by means of suitable agentic service facilities and tools remains an important research topic for AMServ. It is critical to preserve user autonomy, when their perceptual environment is continuously curated by AMServ.

\section{Conclusion}

% The new technologies, e.g., GenAI, LLMs, agentic methods and the metaverse, bring service computing and service engineering into a new stage, with many significant challenges. Smart services, agentic services and metaverse services become the new forms of services supported by GenAI. As the cross point of these new technologies, the Agent-as-a-Service (AaaS) in the metaverse environment (Meta-AaaS) is an emerging and exciting paradigm of agentic services tailored to metaverse environments.

% Based on analysis and comparison between the Meta-AaaS with traditional approaches of agents and service computing, this paper discusses the basic concepts, characteristics, principles, architecture, roles, and advantages of the Meta-AaaS, and presents the typical applications of the Meta-AaaS. The Meta-AaaS represents an important extension of the existing service models. By embedding goal-directed agents as service providers within immersive cyber-physical environments, a possible shift becomes possible from interface-driven to intent-driven service computing. The open research topics identified in this paper provide a preliminary research agenda for the vibrant field. As GenAI capabilities continue to mature and metaverse infrastructures evolve, Meta-AaaS will become an important research field for next-generation services, which will lead the development of the emerging virtual service industries and economy in the future digital and intelligent ecosystems.

The new technologies, e.g., generative AI (GenAI), large language models (LLMs), agentic methods and the metaverse, bring service computing and service engineering into an exciting new stage, yet with many significant challenges. Smart services, agentic services, and metaverse services have become the new forms of services supported by GenAI. At the intersection of these new technologies, the Agent-as-a-Service (AaaS) in the metaverse environment (Meta-AaaS) is an emerging and exciting paradigm of the Agentic Metaverse Services (AMServs), the agentic services tailored to metaverse environments.

Based on the analysis and comparison between the Meta-AaaS and traditional approaches of agents and service computing, this paper discusses the basic concepts, characteristics, principles, architecture, roles, and advantages of AMServ and Meta-AaaS, and presents the typical applications of AMServ and Meta-AaaS. AMServ and Meta-AaaS represent the important extension of the existing service models. By embedding goal-directed agents as service providers in immersive cyber-physical environments, a shift becomes possible from interface-driven to intent-driven service computing. The open research topics identified in this paper provide a preliminary research agenda for this vibrant field. As the GenAI capabilities continue to mature and the metaverse infrastructures evolve, AMServ and  Meta-AaaS will become an important research field for next-generation services, which will lead to the development of the emerging virtual service industries and economy in the future digital and intelligent ecosystems.

% 末尾引入参考文献
\bibliographystyle{IEEEtran}
\bibliography{references}

\end{document}